# Linear scanning ATR-FTIR for chemical mapping and high-throughput studies of *Pseudomonas* sp. biofilms in microfluidic channels

Mohammad Pousti,[a] Maxime Joly,[b] Patrice Roberge,[b] Mehran Abbaszadeh Amirdehi,[a] Andre Bégin-Drolet,[b] Jesse Greener*[a,c]

[a] Département de chimie, Faculté des sciences et de génie, Université Laval, Québec (QC) G1V 0A6, Canada

[b] Département de génie mécanique, Faculté des sciences et de génie, Université Laval, Québec (QC) G1V 0A6, Canada

[c] CHU de Quebec Research Centre, Laval University, 10 rue de l'Espinay, Québec (QC) G1L 3L5, Canada

**ABSTRACT:** A fully automated linear scanning attenuated total reflection (ATR) accessory is presented for Fourier transform infrared (FTIR) spectroscopy. The approach is based on the accurate displacement of a multi-bounce ATR crystal relative to a stationary infrared beam. To ensure accurate positioning and to provide a second sample characterization mode, a custom-built microscope was integrated into the system and the computerized work flow. Custom software includes automated control and measurement routines with a straightforward user interface for selecting parameters and monitoring experimental progress. This cost-effective modular system can be implemented on any research-grade spectrometer with a standard sample compartment for new bioanalytical chemistry studies. The system was validated and optimized for use with microfluidic flow cells containing growing *Pseudomonas* sp. bacterial biofilms. The complementarity among the scan positioning accuracy, measurement spatial resolution and the microchannel dimensions paves the way for parallel biological assays with real-time control over environmental parameters and minimal manual labor. By rotating the channel orientation relative to the beam path, the system could also be used for acquisition of linear biochemical maps and stitched microscope images along the channel length.

## Introduction

The field of bioanalytical chemistry is currently undergoing rapid development. Trends toward more precise characterization, high-throughput analysis and greater levels of automation collectively offer the promise of turnkey systems that can deliver deeper insights into living biological systems. Commercialized examples include plate readers, flow cytometers and patch clamp systems. Additionally, standard analytical chemistry instrumentation such as NMR spectroscopy and mass-spectrometry can be modified or used as-is in longitudinal biological studies.[1,2] A key point in such applications is fluid and sample handling. Manual approaches such as pipetting or use of pin tools can be accurate but are time-consuming and laborious. Robotic samplers, auto-injectors and bulk liquid dispensers can reduce manual labor and limit human error, but drawbacks include high costs, large spatial footprints and challenges in coupling to typical analytical chemistry tools. In any case, the hydrodynamic environment is usually not controlled, despite its fundamental importance in the growth and activity of microorganisms. Therefore, the need exists to integrate versatile multi-modal characterization with real-time control over solution-phase parameters for study of living biological samples.

Research into biofilms is accelerating due to their roles in the environment, human health, industrial biofouling and their potential as biocatalysts.[3,4] The study of bacterial biofilms is expected to benefit from new approaches to bioanalytical chemistry because their chemical and structural properties are complex, heterogeneous and undergo changes in time. Biofilms consist of sessile bacteria surrounded by a self-produced extracellular polymeric matrix (EPM) composed of nucleic acids, proteins and polysaccharides. The biofilm and its EPM are highly sensitive to environmental conditions. In particular, the physiochemical properties of the liquid-phase conditions can have profound effects on growth rate, structure, pH, molecular mass transport and metabolic activity.[5-9] Commercially available flow cells are commonly used to culture and observe adherent microbiological species and their biofilms. These devices are typically millifluidic and glass-based and are often used in a three-channel flow cell setup for replicates or limited assays. Their planar transparent geometries make them compatible with selected optical techniques, primarily transmission, fluorescence or confocal laser scanning microscopy, but examples of sensor integration for broadband chemical analysis are limited. In addition, despite miniaturization, large volumes of required nutrient solution can inhibit long-duration experiments, especially those conducted over wide ranges of flow velocities.

Microfluidic flow cells operate at the submillimeter scale, resulting in reduced material consumption, lower cost, strictly laminar flow conditions, and greater potential for parallelization. The major practical advantages of such small channels include applicability to long term-studies, even at elevated shear stresses,[10] as well as better control over key liquid-phase parameters such as chemical concentrations, molecular flux, fluid velocity and shear stress.[11,12] Finally, diverse cost-effective fabrication techniques offer greater design flexibility for custom flow channel geometries and generation of secondary flow patterns.[13-15]

To take full advantage of the control supplied by microfluidic systems, measurements should be acquired *in situ*. Optical microscopy has been the historically predominant characterization tool used in microfluidic flow cells, enabling measurements of total biofilm quantity,[16] structural heterogeneity,[17] and dynamic processes with high temporal and spatial resolution. However, until now, chemical-based measurements of biofilm properties have been limited due to technical hurdles in sensor integration within confined microchannels. Nevertheless, flexibility in microfabrication techniques and the range of applicable materials have begun to yield an impressive range of analytical approaches for studying biofilms on a chip, including magnetic resonance imaging, electrochemistry, Raman spectroscopy and infrared spectroscopy.[18,19] Among these methods, surface-sensitive attenuated total reflection Fourier transform infrared spectroscopy (ATR-FTIR) is particularly compatible due to biofilm preference for surface attachment. In addition, evanescent light from ATR-FTIR can be applied directly to the biofilm, thereby avoiding attenuated signals due to strong water absorption. However, long-range spatially resolved ATR-FTIR acquisition within an enclosed microchannel is less practical. One ground-breaking approach involved adherence of a microchannel to a large-area ATR crystal and irradiation with a defocused IR beam to generate FTIR chemical maps inside microfluidic channels.[20] The approach benefitted from its versatility, enabling two dimensional sensing with good spatial resolution. However, the setup is highly specialized and the array detector is particularly expensive. Moreover, this technique interrogates the entire ATR crystal surface, despite the fact that microchannels only overlap with a small fraction of the crystal surface area. A large fraction of the infrared source light is thus spent probing non-essential locations, impeding applications to weakly absorbing systems such as hydrated biological samples.

In this work, we demonstrate a method for spatially resolved ATR-FTIR measurements in microchannels that uses precision movement of the device relative to the IR beam from any standard research-grade spectrometer. By sacrificing two-dimensional measurements in favor of spectra collected along a linear path, the approach achieves both good sensitivity and cost-efficiency, while maintaining imaging dimensionality that complements straight microchannel geometries. A fully integrated optical microscope offers parallel image acquisition and feedback for scan position accuracy. Custom software enhances system usability and automates all aspects of long-term experiments. The system is demonstrated in assay mode and linear mapping mode for growth experiments on *Pseudomonas* sp. CT07 biofilms.

## Materials, equipment and methodology

**Infrared spectroscopy:** Infrared measurements were conducted using a research-grade FTIR spectrometer (iS50, Thermo Fisher Scientific, MA, USA). System functionality was also demonstrated on two older FTIR spectrometers (Magna 560, Nicolet, USA and M120, Bomem Inc., Quebec, Canada) with a liquid nitrogen-cooled narrow-band MCT detector. For all measurements, the number of scans was 64 and the spectral resolution was 4 cm$^{-1}$. A custom-built ATR accessory with a germanium (Ge) multibounce crystal is described below. Data acquisition and spectral processing were performed with software (OMNIC v5.2, Nicolet, USA and GRAMS/AI v8.0, Thermo Fisher, USA, respectively). Spectral overlap from water vapor, liquid water and PDMS were eliminated using the appropriate background spectra.

**Microscope imaging:** The device was fully transparent and contained a reflective Ge crystal that enabled simultaneous characterization by optical microscopy. Microscope images were acquired using a 12-bit, 5-megapixel camera (EQ-5012M MONO, Edmund Optics, Canada) attached to a variable-zoom (2.5x to 10x) imaging lens (VZM 1000i, Edmund Optics, Canada). The long working distance (35 ± 1 mm) of the lens supplied sufficient space between the microscope optics and the microfluidic device to prevent any collisions with connective tubing or the ATR accessory during the scanning process. Illumination was supplied through a fiber-optic ring light system (Ring Light Guide, 54-175, Edmund Optics, Canada). Illumination was delivered off-axis (68°), resulting in biological samples that appeared bright against a black background, similar to darkfield illumination, for more sensitive detection of early biofilms. The illumination system was connected to a portable light source (Fiber-lite, Dolan-Jenner Industries, MA, USA) via a home-built optical coupler. Fine positioning of the microscope assembly was implemented using a three-dimension rack-and-pinion stage (55-621, Edmund Optics, Canada) connected via c-mount. The assembly was mounted onto the spectrometer base-plate such that its imaging area was fixed, and thus the scanning ATR stage and attached microfluidic device could be positioned as required. Camera control was achieved with open-source software (Micromanager 1.4, Open Imaging, USA) and using hardware specific libraries (µEye, IDS Imaging Development Systems GmbH). Image analysis was performed using ImageJ v1.51 (National Institute of Health, USA).

**Microfluidic device fabrication with embedded ATR crystal probe surface:** A mold for polydimethylsiloxane (PDMS) casting was fabricated using a dry (negative) photoresist film (Photopolymer film 55 µm, Mungolox, Germany) adhered to a glass slide with laminator (304 Model, Fortex, UK). To achieve channel heights of 110 µm, two layers of films with 55 µm thickness were laminated onto the slide. The channels were defined using a photomask printed with a photoplotter (FPS 250000, Fortex). All masks were designed using Autocad software and printed on a transparent acetate sheet to create the desired mask. The mask allowed selective cross-linking in the photoresist under exposure to UV light (UV-AY315, Fortex, UK). Unexposed portions of the photoresist were removed using the supplied developer solution, which consisted primarily of sodium bicarbonate. All channel heights were 110 µm, whereas the length and width were changed based on the experiment. See Supporting Information for the three main channel designs used in this work. They included (i) a single "false" channel with unsealed channel ends to limit PDMS signal contamination during resolution measurements, (ii) a six channel device for parallel measurements and assays and (iii) a single channel for linear mapping. In the latter two, PDMS in the IR beam path (at the channel ends and side-walls, respectively) caused residual $CH_3$ signal in the acquired spectra. Microfluidic devices were fabricated by casting of PDMS and cross-linker (Sylgard184, Dow Corning, Canada) at a 10:1 ratio onto the mold. After 4 hours at 80 °C, the hardened PDMS was removed from the mold. The open PDMS microchannel was sealed by a 24-bounce Ge ATR crystal (50 mm × 20 mm × 3 mm, 45° parallelogram) after air plasma activation (PCD-001 Harrick Plasma, Ithaca, USA) at 650 torr for 90 s. The ATR crystal was not treated in the plasma cleaner at the same time as the PDMS device to avoid surface contamination that has recently been linked to changes in the biofilm growth rates in microfluidic experiments for the same bacterial species as used here.[21] A plastic top plate was used to reinforce the leak-proof interface between the PDMS device and the Ge ATR crystal. After the experiments were completed, the microfluidic device and ATR crystal were gently separated by hand. Trace amounts of strongly adhered PDMS on the crystal were removed by prolonged submersion in an acetone solution, followed by submersion in a soap solution and light polishing with a cotton swab.

**Spectral data treatment:** The system had the potential to generate a large amount of data during the experiments which typically lasted two to three days. Systematic analysis of IR data was conducted using a combination of pre-existing functions (Grams/AI software environment) and custom macros for background subtractions (primarily $H_2O$ (l) and PDMS and as well $CO_2$ (g), $H_2O$ (g), offset corrections and location of peak heights.

**Fluidic interface and control:** Inlets and outlets were connected using metal elbow capillaries that tightly fit into the punched holes without the need for epoxy. Syringe pumps (PHD 2000, Harvard Apparatus, Holliston, MA, USA) were used to inject liquids into the microchannels. Connective tubing made of perfluoroalkoxy (PFA) with an outer diameter of 1.6 mm (U-1148, IDEX, WA, USA) was connected to 60 mL syringes (BD Scientific, NJ, USA) via a connector assembly (P-200x, P-658, IDEX, WA, USA). All liquids were degassed to prevent bubbles that could otherwise cause local alterations to biofilm formation[22] or enhanced shear forces that could disrupt biofilm attachment.[23, 24]

**Biofilm cultivation and inoculation:** A pre-culture of planktonic *Pseudomonas* sp. CTO7 was used as inoculum for biofilm formation. The suspended culture inoculum was obtained by shaking planktonic bacteria in 5 mL of 5 mM growth media in an incubator for 18 h at 30 °C. The growth media was a modified AB type consisting of 1.51 mM $(NH_4)_2SO_4$, 3.37 mM $Na_2HPO_4$, 2.20 mM $KH_2PO_4$, 179 mM NaCl, 0.1 mM $MgCl_2 \cdot 6H_2O$, 0.01 mM $CaCl_2 \cdot 2H_2O$ and 0.001 mM $FeCl_3$ with 10 mM Na-citrate$\cdot 6H_2O$ as the sole carbon source. Prior to inoculation, all fluidic components were disinfected with 70% ethanol for three hours and subsequently rinsed for one hour with distilled water. Inoculum was injected manually into the microchannel by piercing the device with a high-gauge syringe needle. The inoculant solution was left in place for 2 hours, followed by a flow of sterile growth medium via syringe pump through pre-connected tubing. Characterization began less than one hour after nutrient solution began flowing. All experiments were conducted in a temperature controlled laboratory (22 ± 1 °C). As PDMS is well-known to be highly permeable to small gas molecules, the biofilms were well oxygenated and buildup of by-products, such as $CO_2$ was limited.

## Results

**System development:** In addition to the customized microfluidic device with an embedded ATR element, the complete analytical system included four main components: (i) a custom-built horizontal ATR accessory fixed to a scanning stage, (ii) a custom-built optical microscopy system, (iii) a FTIR spectrometer and (iv) customized software for control over (i)-(iii) and data management. The portable linear scanning ATR accessory was formed by components (i) and (ii). A CAD-generated schematic of the accessory is shown in Figure 1a. The scanning ATR stage was based on dis-placement of a standard x-y positioning stage with manual micrometer-based motion control. The linear scanning ATR accessory was placed in the spectrometer sample compartment and adjusted to match the z-height of the IR source beam. The y-direction micrometer was manually adjusted to center the focal point of the converging IR beam on the mid-point of the ATR crystal sensing surface and was not subsequently moved. A stepper motor (NEMA23 hybrid step bipolar, Trinamic, Germany) with a repositioning accuracy of $6.6 \times 10^{-4}$ deg/rotation was coupled to the x-direction micrometer using a 1/8-inch-wide MXL sized timing belt (104MXL012, #7887k23, McMaster Carr, USA) and corresponding drive pulleys on the motor and micrometer (diameter ratio of 0.734). This setup eliminated any measureable repositioning offset error due to slippage during the course of a typical experiment. The stepper motor displaced the scanning ATR stage perpendicular to the IR beam with accuracy at the sub-micron scale. The scanning ATR stage was positioned such that the overall x-direction scan range (15 mm) allowed the IR beam to be displaced applied across the ATR crystal beveled edge. As discussed in the Supporting Information (Section 5), scan distances were limited to an 11 mm segment of the Ge ATR crystal. The IR beam and microscope field of view were both static and co-localized such that movement of the scanning ATR stage enabled overlapping measurements of both modalities at different positions on the microfluidic device.

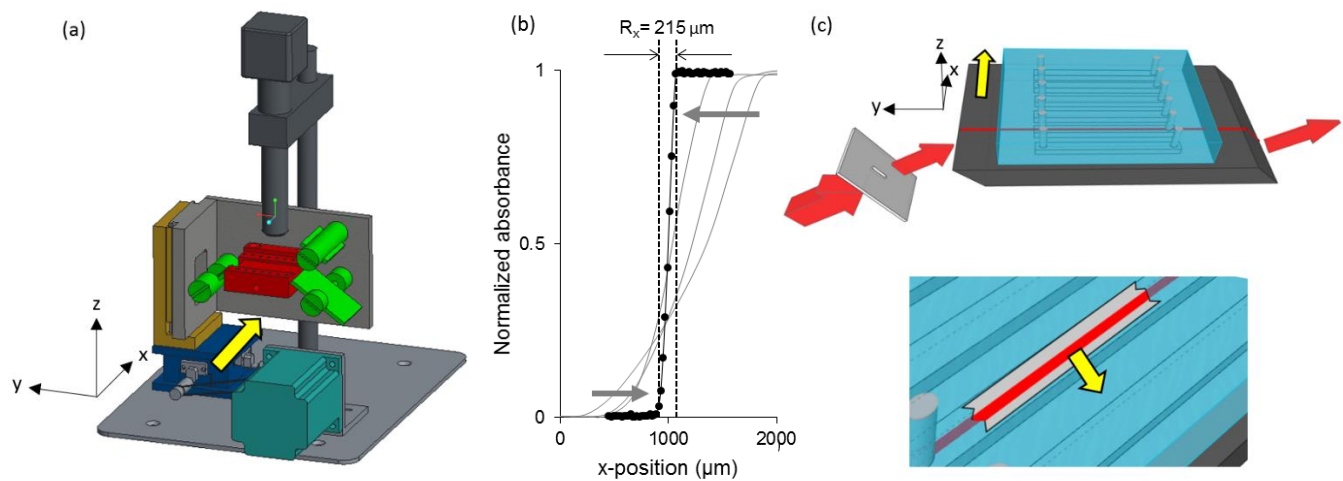

Figure 1. (a) CAD schematic of the portable scanning ATR accessory with components identified by color. The base plate support (grey) and movable stage (blue) are controlled by micrometer positioners (grey) coupled to a stepper motor (dark green) via a timing belt (black). An optical holder (brown) immobilizes the moving stage and a multi-bounce ATR accessory with light guides (green) and ATR crystal holder (red) against the movable stage. The yellow arrow indicates the scan direction. (b) Data showing the increase in absorbance asymmetric stretch of $CH_3$, 2960 cm$^{-1}$ in PDMS during scanning from the ATR-air interface toward the ATR-PDMS interface at the microchannel wall using different internal apertures ranging in diameter from 2.5 to 0.75 mm. Transition from microchannel to PDMS wall showing trends toward smaller rise distances as the internal aperture size is decreased (grey arrows). Black points display the highest resolution achievable ($R_x$ = 215 μm). (c) Cartoon of a six-channel PDMS microfluidic device (blue) with embedded ATR crystal (dark grey). IR light (red) is reduced with an aperture (grey) before hitting the beveled edge (angle of incidence 90) and then traveling through the Ge ATR crystal at a 45 degrees relative to the sensing surface. The yellow arrow indicates the same scan direction shown in (a). Inset shows a zoomed view on one channel with a breakaway view within displaying the IR beam path. The yellow arrow indicates the relative displacement of the IR probe light toward the channel wall during scanning.

Computer control over the stepper motor and camera operation was first achieved via separate programs in Python language running on a stepper control board (TMCM-1160-TMCL Trinamic, Germany) and via using open access software with camera-specific libraries (described above). Following successful development, these components were subsequently merged into a LabVIEW program where all necessary functionality and control were integrated. The software was compiled to be run as an executable file on any computer. The graphical user interface allows the system operator to easily set measurement parameters. For microscope imaging these included camera brightness, pixel clock, frame rate, acquisition time and for FTIR spectroscopy these included sampling rate, number of scan co-additions and spectral resolution. The program also controls the ATR crystal displacement and coordinated with experimental parameters based on user preferences and microchannel geometry (cycle time, displacement velocity, sampling locations and dwell times at each location, number of channels, spacing between channels, etc.) These parameters are communicated to the FTIR software (Omnic v5.2, Nicolet Instrument Corp., WI). The software also included a live feed from the camera for live monitoring of the experiment. Adjustments to (8-bit) pixel brightness could be optimized manually using a high/low display to highlight empty pixels (with value=0) and saturated pixels (with value=255) in real-time. A second automated gain setting mode is also available (Supporting Information). Further details regarding the software design and user interface are available in the Supporting Information.

**Determination of spatial resolution and system validation:** The first validation experiment determined the effective spatial resolution of the IR measurements in the x-direction, $R_x$ (Figure 1b). This measurement was performed using an ATR microfluidic chip containing a single-channel device with width w = 3 mm and length L = 1 cm (see Supporting Information for more details). The scanning ATR accessory was positioned such that IR beam was located in the middle of the microchannel, which avoided detection of PDMS from the side-walls (via $CH_3$ vibrations at 2960 cm$^{-1}$). This marked the x = 0 position in Figure 1b. Data acquisition continued as the microchannel wall was moved closer to the IR beam by increasing the x-position. At a certain point the $CH_3$ signal began to increase. The x-position that resulted in the highest $CH_3$ absorbance and maintained that value during further displacements was assumed to be located fully outside the channel. Measurements of the displacement distance required to transition the absorbance from 0 to its maximum value determined the effective beam width, $R_x$. The scan path was repeatedly retraced, each time with a smaller aperture being applied. As the internal aperture size was decreased, the slope in the low-to-high signal transition increased, representing an improvement to spatial resolution. The highest spatial resolution of $R_x$ = 215 µm was measured using the smallest aperture. Because the internal aperture is independent of the presented accessory, these results are expected to change from spectrometer to spectrometer. For example on a different FTIR spectrometer that did not contain an internal aperture, a similar resolution was achieved (350 µm) using two home-built external apertures located before and after the scanning ATR accessory presented in this paper. Hence, tailoring of the proposed accessory to different spectrometers, new and old, is possible. While the use of internal or external apertures reduces IR excitation beam intensity (see Supporting Information for more details), measurements using a standard MCT detector readily enabled the acquisition of data with good signal to noise for biofilm samples which are concentrated at the ATR sensing surface. However, the loss of signal may impede sensing of dilute samples in other applications. For example, the improved spatial resolution demonstrated in Figure 1b, which was achieved via a reduction in aperture diameter from 2.5 to 0.75 mm would result in an increase to the limits in the minimum detectible concentration by about three times based on the associated loss of detector signal. The signal intensity and spatial resolution might be further improved in future developments with the addition of one or more external apertures or lenses directly on the portable scanning ATR accessory. In this proof-of-principle demonstration, we increased the aperture size to above the minimum, resulting in $R_x$ in the range of 0.6 to 0.7 mm for the channel used in the remainder of this study. This process ensured that the peak absorbance occurred directly in the middle of the channel (w = 1.5 mm) for up to six parallel channels on a single ATR crystal (Figure 1c).

For the next part of the work, a microfluidic ATR device with 6 parallel rectangular microchannels attached to a Ge ATR crystal was used. Their long axes (L = 2 cm) were oriented parallel to the IR probe beam (y-direction) such that the entire length of each channel could be probed independently (see Supporting Information). A stationary camera was trained on the sample from the top-side and adjusted such that the field of view overlapped with the beam IR beam position. Parallel spectroscopy and image acquisition was conducted as the scanning ATR stage was displaced such that the IR beam was swept across the entire device. Figure 2a shows a stitched image from in-line microscopy focusing on the five PDMS walls separating the six channels along a strip of the microfluidic device. This image was compared with the $CH_3$ absorption peak from PDMS, which oscillated as the IR beam encountered and passed by the open microchannels (Figure 1b). To ensure that the microscope field of view was centered on the IR beam, we integrated the image pixels that overlapped with the PDMS material in a window matching the IR beam width. These results show in-phase oscillations from the two characterization modalities, thus demonstrating their alignment. In addition, the rise and fall of the curve shapes verified $R_x$ of 0.6 mm (Figure 2b). As explained in the sub-section "Microfluidic device fabrication with embedded ATR crystal probe surface" and as shown in Table S1 in Supporting Information, the non-zero minima measured by ATR-FTIR was the result of the IR beam interaction with the PDMS at the beginning and end of the channel, which was required for proper sealing against the ATR crystal. Figure 2 highlights the IR beam path relative to the optical image in three different positions of beam placement: (1) inside the channel, (2) straddling the channel/PDMS interface and (3) between two channels. We note that in the case of (1), the $CH_3$ signal from PDMS was nearly, but not totally, constant for small displacements away from the channel center position. This is likely due to a slight Gaussian beam shape or the converging/diverging beam profile away from the focal point, which can result in a small amount of light reaching the walls at the aperture setting used in this experiment. Next we validated that the contents of each channel could be measured separately without any contamination from neighboring channels.

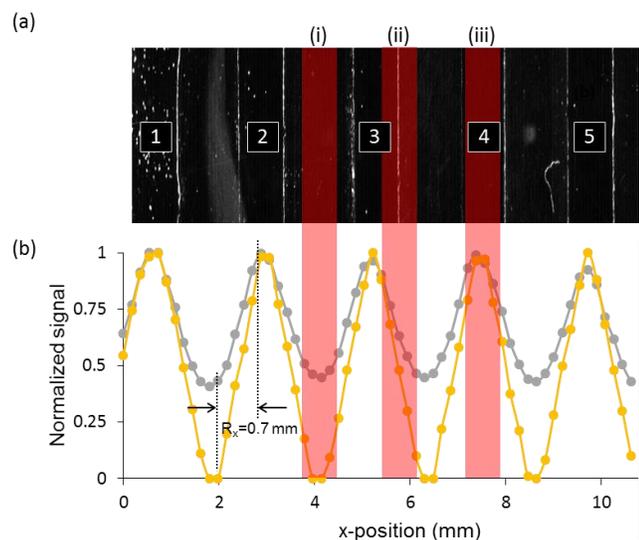

Figure 2. (a) Stitched image showing a portion of a parallel six-channel device with five enumerated separation walls. Scale bar is 1 mm. (b) Normalized IR absorbance of PDMS using IR (grey) and optical image analyses (yellow) at the indicated x-position (x-scale is applicable to (a) as well). Red lines connecting the image to 1D intensity plot are placed at the (i) channel center, (ii) straddling the PDMS channel interface and (iii) between channels. $R_x$ is shown for the channel to wall 2 transition.

To validate the independence of spectroscopic measurements between channels, a six-channel device with channel orientation parallel to the IR beam (Figure 3a) was filled with $H_2O$ and $D_2O$ liquids in alternating channels. Figure 3b shows the distinctive vibrational bands for $H_2O$ (OH stretching and bending at 3370 and 1640 cm$^{-1}$, respectively) and $D_2O$ (OD stretching at 2490 cm$^{-1}$) spectra collected from the channel midpoints of two adjacent channels, and the $CH_3$ bands (1260 cm$^{-1}$) from pure PDMS can be observed between them. As these PDMS bands were narrow and localized, the proper choice of characteristic analyte bands could eliminate overlap. It should be noted that a major doublet is present for PDMS in the low frequency range of 1100–1000 cm$^{-1}$ (data not shown). Except for a residual signal from $CH_3$ in both channels due to PDMS at the beginning and end of each channel, the $D_2O$ and $H_2O$ signals were pure and completely unmixed. To verify the ability to collect independent measurements from all channels, a full scan along the x-direction was conducted with step size $\Delta x = 180$ µm (Figure 3c). The results showed oscillating intensity of the characteristic bands for each liquid with an out-of-phase oscillation in the $CH_3$ absorbance. Similar plots were acquired for microfluidic devices with 4 and 5 channels (Supporting Information).

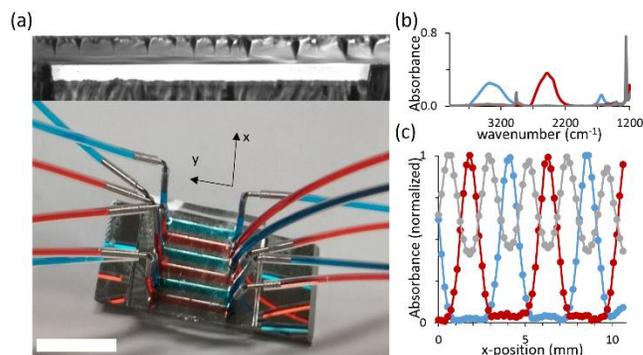

Figure 3. (a) The six-channel device on top of the Ge ATR crystal used for parallel assays in this work. Connections between the inlet and outlet tubing were made with metal elbow joints. Channels are filled with alternating colored liquids for visualization. Scale bar is 2 cm. Inset (top) shows the cross-section of one channel attached to a glass surface with length = 20 mm, height = 110 µm and width = 1.5 mm. The x-y axes define the coordinate system used in this paper. (b) Spectra acquired from (i) channel containing $D_2O$ (red), (ii) channel containing $H_2O$ (blue) and (iii) PDMS spacer between two channels (grey). All spectra were acquired with 64 scans using a background from pure air. (c) Normalized absorbance values of characteristic peaks for $H_2O$ (3370 cm$^{-1}$), $D_2O$ (2490 cm$^{-1}$) and PDMS (1260 cm$^{-1}$) at different x-positions across the ATR crystal.

**Measurements on growing biofilms:** To demonstrate the system, continuous measurements of a growing biofilm sample in a single channel was conducted. The measurements started after inoculation for two hours, followed by a continuous flow of an AB nutrient solution containing 10 mM citrate at $Q = 0.2$ mL·h$^{-1}$. The resulting average flow velocity of 0.61 mm·s$^{-1}$ resulted in a calculated on-chip transit time of $t = 33$ s. Figure 4a shows a typical spectrum of a mature biofilm (50 h after inoculation) after spectral subtraction of PDMS and water. The major biofilm bands included amide I (C=O stretch coupled with N–H bend) and amide II (C–N stretch coupled with N–H bend), which are characteristic of proteins (1640 cm$^{-1}$ and 1540 cm$^{-1}$). Secondary absorbance peaks were observed from $CH_2$ vibrations in lipids (2920, 2850 and 1450 cm$^{-1}$),[18,24,25,26] from COO$^-$ in amino acids (1400 cm$^{-1}$) and from phosphate-containing phospholipids and phosphodiesters nucleotides (1085 and 1235 cm$^{-1}$). The peak at 1085 cm$^{-1}$ is complex, with high and low frequency shoulders located at 1120 and 1050 cm$^{-1}$ arising from contributions by DNA nucleotides and polysaccharides, respectively. The amide II band is typically selected for longitudinal analysis due to its strength and isolation from other bands such as the water band, which overlaps with amide I. In the current case, PDMS can interfere with most biofilm peaks at frequencies less than 1235 cm$^{-1}$. Therefore, for this proof-of-principle study, we analyzed the amide II band in the remainder of this work. Figure 4b shows the increase in the amide I and amide II absorbance peaks in proteins during the first 50 h of growth. A small band between these two peaks originates from AB citrate molecules, which appear to accumulate in time. Figure 4c shows the imaging results focusing on a portion of one channel of the parallel channels during ATR-FTIR measurements on the same channel. The red box in the first image shows the approximate path of the IR beam down the center of the channel. Figure 4d shows the average intensity of the amide II band and the average pixel intensity along the IR beam path with time. Biofilm growth near the side-walls was not included in either the IR measurements or the calculated pixel intensity. The exclusion of the channel walls in the measurement region avoided local perturbations in biofilm production, especially for low-aspect-ratio channels such as those used in this work.[27,28] The results of the growth experiment are presented on a normalized scale for comparison. The initial increase in the amide II signal was more rapid than that of the pixel intensity from optical imaging, which shows a lag phase for approximately 10 h. This result demonstrates the higher sensitivity of ATR-FTIR spectroscopy to initial growth relative to the microscope imaging.

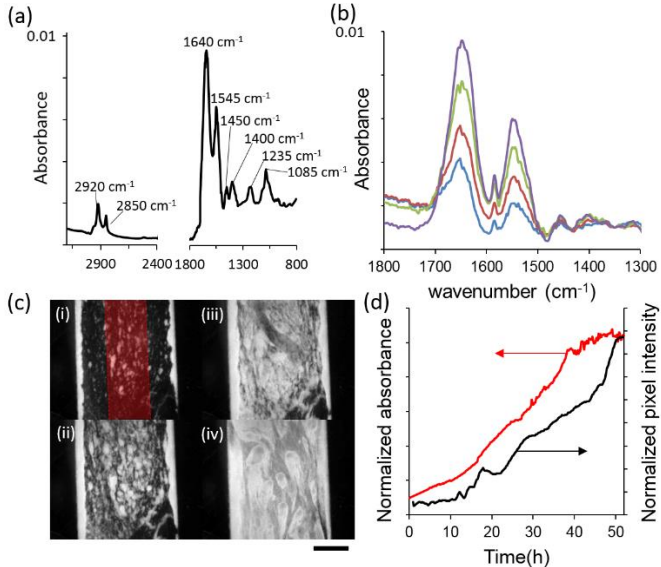

Figure 4. (a) IR spectra of a *Pseudomonas* sp. biofilm (50 h) collected from a single channel. The highlighted peaks result from $CH_2$ (2920, 2850, and 1450 $cm^{-1}$), amide I (1640 $cm^{-1}$) amide II (1545 $cm^{-1}$), amino acids (1400 $cm^{-1}$), and phosphate-containing groups (1085 and 1235 $cm^{-1}$). (b) Growth of the amide I and II bands in the range of 1300–1800 $cm^{-1}$ at 20 (blue), 25 (red), 35 (green) and 50 h (purple). (c) Optical micrographs of a portion of the same channel measured in (b) at the same growth times. Scale bar is 500 µm. Image window is 1.7 × 1.5 mm. (d) Average absorbance values (red) and average optical intensity (black) versus time. Flow rate $Q = 0.2$ mL·$h^{-1}$. All spectra are the results of spectral subtraction of pure PDMS and water.

**System repeatability in bacterial biofilm measurements:** We also address a fundamental challenge in bioanalytical chemistry, namely, reduction of the "noise" floor arising from the large variations of living biofilm properties, which can arise from slight differences in biological and ambient conditions that cannot be completely controlled by the experimental platform during sequential measurements. The ability of parallel microchannel experiments to reduce batch-to-batch variations of inocula, nutrient solutions and ambient conditions such an approach was tested. This was accomplished by comparing biofilm growth kinetics from parallel on-chip experiments with those from sequential experiments conducted in identical microfluidic channels.

In the parallel measurements, fluid was supplied separately to each channel with a different syringe to avoid the possibility of flow redistribution and cross-contamination. The averaged time series of protein absorbance (amide II band) and pixel intensity measurements are shown in Figure 5a, with the standard deviation in each represented by error bands. These data were compared with the results acquired from six sequential experiments shown in Figure 5b, which were conducted on biofilms that were inoculated on different days from different inoculum solutions but grown under the same experimental conditions. A comparison of Figures 5a and 5b shows that the measurement error bands were significantly larger for rep-licate measurements conducted sequentially than for those conducted in parallel. Figure 5c quantifies this observation by plotting the repeatability enhancement (RE) achieved by conducting six replicate measurements in parallel compared with those performed sequentially. The RE, which compares the standard deviations of replicate experiments conducted sequentially ($STD_{seq}$) and in parallel ($STD_{parallel}$), was calculated for both optical intensity and amide II absorbance values using Eqn. 1:

$$RE = STD_{seq} / STD_{parallel} \qquad (1)$$

The RE values were generally >1, indicating that parallel measurements improved repeatability. The only exception occurred during the first 20 hours when optical microscopy RE was roughly 1, likely due to the lower sensitivity of the technique, which resulted in an extended measured lag phase. In contrast, measurements collected by surface-sensitive ATR-FTIR strongly benefited from parallel measurements in the first hours after inoculation (Figure 5c inset). Both measurements show increased RE during the transition from the lag phase to the rapid growth phase (22 h) and again during the transition from the rapid growth phase to the stationary phase (44 h). Therefore, parallel measurements are generally beneficial, but are especially important during the transitions to new growth phases. Addressing this point is important so that precise quantitative tools such as FTIR spectroscopy can deliver the best results possible.

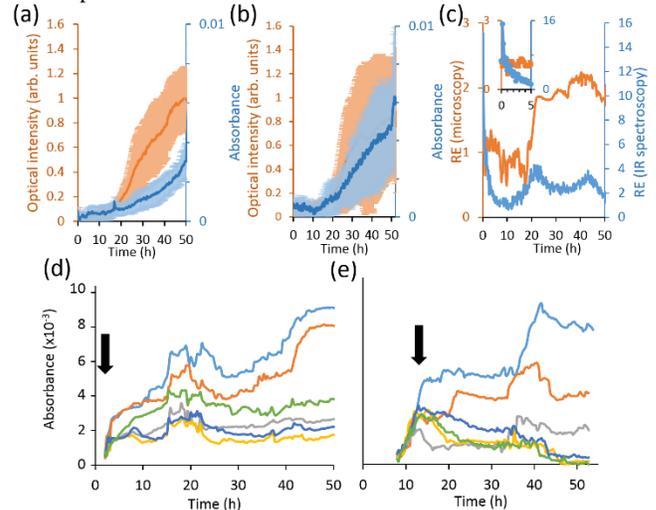

Figure 5. (a) Average intensity of amide II absorption band (blue) and pixel intensity (orange) for 6 sequential *Pseudomonas* sp. biofilm cultures from (i) separate experiments conducted on different days with different inoculum and (ii) experiments conducted in parallel on a six-channel device with the same inoculum. Error bars represent the standard deviation (STD). In all cases, an AB nutrient medium with 10 mM citrate was applied with a flow rate $Q = 0.2$ mL·$h^{-1}$ in channels with the same dimensions (w = 1.5 mm, L = 2 cm, h = 110 µm). (c) Repeatability enhancement (RE) for microscopy and IR spectroscopy. Inset shows RE during the first 5 hours. (d) Growth experiments with varying concentrations of ethanol added to an AB nutrient solution immediately following inoculation with concentrations (w/w%) of 0 (light blue), 1 (orange), 2 (grey), 3 (yellow), 4 (dark blue) and 5 (green). (e) Growth experiments with ethanol concentrations (w/w%) of 0 (light blue), 3 (orange), 5 (grey), 10 (dark blue), 20 (yellow) and 70 (green) added to an AB nutrient solution 14 h following channels inoculation.

To demonstrate the utility of parallel growth channels for conducting assays on replicate biofilm samples, we evaluated the effect of nutrient solutions containing ethanol at different concentrations on pre-inoculated surfaces and nascent biofilms using ATR-FTIR. Ethanol presents efficient anti-bacterial effects on planktonic and pre-biofilm sessile bacteria at concentrations greater than 4 w/w%[29], and higher concentrations for those protected by biofilms.[30] Ethanol at different concentrations was added to the AB nutrient solutions containing 10 mM citrate following inoculation of *Pseudomonas* sp. bacteria (Figures 5d and 5e). Initially, ethanol concentrations of 0–5 w/w% were applied to the biofilm after a 2 h growth period following inoculation. The results show that ethanol acts as an inhibitor, dramatically reducing biofilm protein production for ethanol concentrations greater 2 w/w%, whereas 1 w/w% ethanol had no effect relative to a pure AB nutrient solution. In a second assay experiment, the bacteria were allowed to grow after inoculation for 14 h before application of ethanol to achieve antifouling effects in nascent biofilms. The results from this experiment showed inhibition up to 5 w/w% and decreasing protein intensities with time for ethanol concentrations of 10 w/w% and greater, indicating that biofilms were not only inhibited but were also removed from the ATR crystal surface.

To demonstrate the potential of linear scanning ATR-FTIR as a mapping technique for low-dimensional samples, such as in straight microchannels, the channel was rotated by 90 deg such that its intersection with the IR beam could be displaced along the length of the microchannel using the scanning ATR stage. In this orientation, the scan axis (x-direction) was parallel to the channel and linear spectral maps were generated. Figure 6a shows a schematic of a segment of the ATR crystal featuring a channel with w = 3 mm and L = 1.1 cm placed with its long axis perpendicular to the IR beam. This channel was approximately two times wider than the other channels to enable additional interaction with the IR probe beam crossing its short axis. Only one channel was probed in this work, but the approach could be applied to parallel channels, although the results would be averaged. Figure 6b shows three pairs of images obtained from the local amide II absorbance (top) and the stitched optical image (bottom) at time points of (i) 5 h, (ii) 30 h and (iii) 60 h. Figure 6c shows a space-time map with each horizontal pixel row representing the spatial variation of amide II absorbance in the channel at a specific time during the 65 h experiment. The figure shows that an initially clean channel begin slowly accumulating protein, and then a switch to more rapid protein accumulation occurred after that. The biofilm began to spread progressively downstream until significant amide II signal was detected throughout the channel at 20 h. Figure 6d shows variations in the amide II absorbance values with time at different channel positions. Interestingly, despite the initial rapid increase in protein signal at the upstream position, the protein signal was nearly halted after a small decrease in its peak value around 20 h. In contrast, the amide II signal increased rapidly after 20 h in the mid-channel position and even more rapidly in downstream positions, with the latter reaching the highest recorded value near 30 h before decreasing significantly. After this time, the downstream amide II signal decreased but maintained nearly constant levels in other positions. This decrease in protein signal at the attachment surface was previously observed in surface-sensitive measurements as an indicator of a "restructuring" process.[31,32] During this process, the biofilm contact with the surface becomes reduced and formation of "mushroom structures" ensues. It is interesting to note that the restructuring process does not appear uniform throughout the biofilm, occurring strongly in downstream portions, weakly in upstream portions and not at all in the mid-stream portions. By averaging the results across the entire channel, evidence of restructuring is completely lost. This indicates that linear mapping measurements are critical to better understand heterogeneous microbiological systems in microchannels.

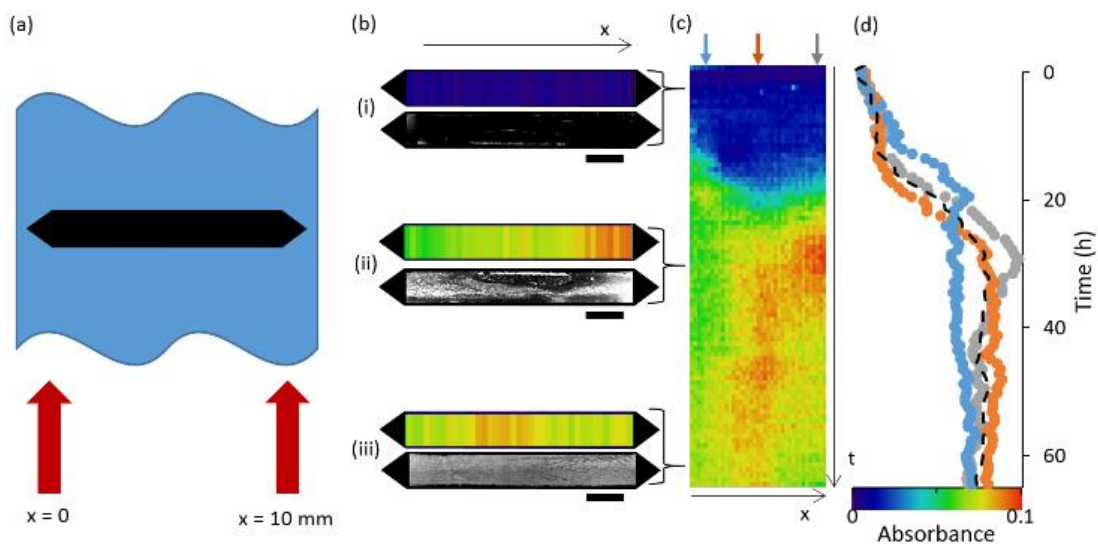

Figure 6. (a) Schematic showing a segment of an ATR crystal (blue) containing the microchannel (black) relative to the IR beam path (red) at different scan positions. (b) Image pairs consisting of a linear map of the amide II absorbance and a stitched optical micrograph at (i) 5 h, (ii) 30 h, and (iii) 60 h, as indicated in (c). Scale bars are 3 mm. (c). Space-time image showing the variation of amide II absorbance along the channel length (horizontal) at different times (vertical). Arrow indicates the scan direction along the x-axis and the direction of time along the vertical axis. Colored arrows indicate x-positions at 0.8 mm (blue), 5.5 mm (orange), and 9 mm (grey) of plots in (d). (d) Time-dependent amide II absorbance at the x-positions marked by colored arrows in (c). The black dotted line is the average amide II absorbance at all positions throughout the channel. The time axis in (d) is matched to that in (c), and the color bar intensities are matched to the absorbance on the horizontal axis. The color bar defines the absorbance values for (b), (c) and (d).

## Discussion

Unlike chemical measurements that are fully repeatable under identical experimental conditions, it is difficult to obtain the same results from repeated measurements of living biological systems because of their natural propensity for heterogeneity and diversity.[32-33] These characteristics are due to intrinsic biological processes related to complex inter- and intracellular mechanisms and system feedback. These factors can both amplify sensitivity to, and obscure the role of, extrinsic physiochemical variables.[34-38] Therefore, even slight differences in experimental parameters can have a large effect on the measured properties, especially in multicellular systems such as biofilms. Even with accurate analytical chemistry tools such as FTIR, bioanalytical studies can suffer from poor measurement repeatability, which strongly limits reliable analysis and transferability to other studies. It has been shown elsewhere that that the combination of microfluidics with *in situ* measurements is one way to overcome these problems. However, in this work, we show that even in a microchannel, where many physiochemical conditions can be tightly controlled, certain challenges remain. For example, trends and local behavior can be obscured because of slight batch-to-batch differences in inocula and nutrient solutions[19,39] or unintended variations in device fabrication or operation, as we have shown recently.[21,22] To this end, the current work seeks to extend the microfluidic advantages related to superior control over the applied flow conditions by also exploiting the small channel dimensions for parallel experiments, which can normalize the starting and ambient conditions. The results showed that parallel measurements were especially important during transitions between different growth phases. This observation is unsurprising because it has been previously reported that variability in batch-to-batch experiments largely arises from the differences in the lag phase duration, which can in turn offset the transition times between other growth phases from one experiment to the next.[40] These offsets are significantly reduced for parallel experiments started at the same time and with the same inoculum, thus opening the door to more reliable assays in which the effects of certain experimental parameters (drug molecule concentrations, for example) can be better correlated with their effects. In conjunction with powerful analytical chemistry tools such as FTIR, rich sample characterization with good repeatability can produce more meaningful results that can be transferable to other studies.

Because biofilm properties are also spatially heterogeneous, it is important to be able to disaggregate local trends from the overall average. With the same portable ATR-FTIR accessory, this was achieved by simply re-orienting the microfluidic flow cell atop the ATR crystal. In the resulting linear imaging mode, we demonstrated the ability to follow the growth behavior of different segments within a continuous biofilm to enable observations of the degree of surface restructuring. The major drawback in this case is the relatively low overlap between the IR beam and the channel (e.g., Figure 6a). The resulting reduction in sensitivity affected the secondary absorption bands but did not impede the ability to monitor protein bands, which are standard target bands for most biofilm studies. Still, with longer integration times and wider channels this issue should be solvable.

Next steps should focus on improvements to the overall system. These might include better measurement spatial resolution via the addition of more sophisticated optical elements, which in turn could enable higher channel density. Software control over fluid pumping could be easily be integrated into the computer software routine. Embedded heating elements can also be added to attain improved control over thermal conditions. Finally, another major step forward would be a method that can limit spectral interference between PDMS and the target analyte bands below 1260 cm$^{-1}$. This would increase the power of the technique by extending spectral characterization of a wider range of biochemical groups. Optimizing the current microfluidic design to reduce the amount of out-of-channel PDMS in the IR beam path is one route. But this would only reduce, not eliminate, the problem. Another more promising approach could be the use of a new generation of low-cost etched ATR elements that can be interrogated at different positions along their bottom side corresponding to channel positions only.[41,42]

## Conclusion

This paper demonstrates a versatile bioanalytical chemistry tool with application to the study of bacterial biofilms. Specifically, a portable attenuated total reflection (ATR) spectromicroscopy accessory with standalone control was devised to enable linear scanning for *in situ* mapping measurements and parallel assays in microfluidic channels using any standard Fourier transform infrared (FTIR) spectrometer. The system was based on the controlled displacement of the ATR accessory relative to the FTIR infrared beam to probe different positions within a microfluidic chip. A custom-built microscope was integrated into the hardware system, offering both a second sample characterization mode and real-time feedback for positioning accuracy. A computer system coordinated the ATR positioning with automated data acquisition and management. After validation, this approach was demonstrated on *Pseudomonas* sp. biofilms in both linear mapping and assay modes. The system paves the way for more reliable assay capabilities as well as the ability to disaggregate local properties from the entire biofilm.

## ASSOCIATED CONTENT

**Supporting Information**

System setup, microfluidic device designs, flow chart and graphical user interface, imaging brightness, additional validation experiments.

## AUTHOR INFORMATION

**Corresponding Author**

Jesse Greener, jesse.greener@chm.ulaval.ca, 418 656-2131 x 7157.

## ACKNOWLEDGMENT

This research was supported by funding from the Natural Sciences and Engineering Research Council, Canada. J. G. is the recipient of an Early Researcher Award and an AUDACE grant (high risk, high reward) for study of microbiological systems using microfluidics from the Fonds de recherche du Québec—nature et technologies (FRQNT). The authors wish to thank Molly Gregas for copy edits.